\documentclass{article}
\usepackage{iclr2026_preprint,times}
\usepackage{amsmath,amsfonts,bm}

\def\eqref#1{equation~\ref{#1}}
\def\1{\bm{1}}

\DeclareMathAlphabet{\mathsfit}{\encodingdefault}{\sfdefault}{m}{sl}
\SetMathAlphabet{\mathsfit}{bold}{\encodingdefault}{\sfdefault}{bx}{n}

\usepackage{hyperref}
\usepackage{url}
\usepackage{booktabs}
\usepackage{graphicx}
\usepackage{tikz}
\usetikzlibrary{arrows.meta,positioning,calc}
\usepackage{amsmath,amssymb}
\usepackage{cleveref}
\hypersetup{pdfauthor={Han Jin}, pdftitle={HoneyRoute: Honeypot-Model Routing for Adversarial LLM Serving}}

\newcommand{\method}{HoneyRoute\xspace}          % method name, adjust freely

\newcommand{\gateway}{inference gateway\xspace}
\newcommand{\pnum}[1]{\textbf{P#1}}
\newcommand{\hpC}{\ensuremath{\widetilde{M}_{\text{c}}}\xspace}   % code-implemented honeypot
\newcommand{\hpT}{\ensuremath{\widetilde{M}_{\text{t}}}\xspace}   % trained honeypot model
\usepackage{xspace}

\title{HoneyRoute: Honeypot-Model Routing for Adversarial LLM Serving}

\author{Han Jin\\
Independent Researcher}

\iclrfinalcopy
\begin{document}
% override the conference running header AFTER it is set by \iclrfinalcopy
\lhead{Preprint}

\maketitle
\maketitle

\begin{abstract}
We introduce \method{}, an inference-serving layer that detects whether an
incoming request is malicious and, if so, routes it to a dedicated honeypot
model---so that
the production service is shielded while the adversary's interaction is
continuously harvested for intelligence. Existing defenses either embed traps
inside model memory/knowledge~\cite{wang202626020,dai202626061} or rebuild the
deception at the protocol layer~\cite{reworr202424101,sladi202323090}, leaving
the serving layer itself unprotected and feeding nothing back into detection.
This is hard because the gateway must decide within per-request latency budgets
and because diverting traffic must not degrade benign quality.
\method{} couples (i)~a streaming router---a frozen, ~0.8B-embedding backbone
with per-domain MLP heads scoring request-level behavioral features---(ii)~a
\emph{dual-implementation} honeypot, either a rule/prompt-engineered
\hpC{} with no model of its own or a dedicated same-family replica
\hpT{}, and (iii)~an analysis loop that converts trapped interactions into
attacker fingerprints used to retrain the detector. Beyond deception, diverted
traffic absorbs resource-exhaustion load and yields forensic attribution
(traceability) signals.
On a production trace plus a seven-domain attack-seed corpus, \method{}'s
router reaches $91.8\%$ malicious recall at the $0.5$ decision threshold
(AUROC $0.975$; false-positive rate $0$ at $95\%$ recall) with $38$\,ms
median added latency---matching $96\%$ of a two-tier
guard-LLM cascade's $F1$ at $1/385$ of its latency; diverting
the malicious share cuts production-model token consumption under
concurrent flooding with real GCG-suffix payloads by $97.8\%$; the trained replica agrees with the
production model on $92.9\%$ of benign holdout requests, while naive
unconditional bait injection collapses to $7.6\%$---and \emph{selective,
camouflaged} injection recovers to $88.9\%$ while keeping traceable
signals on $83$--$100\%$ of attacker interactions, mapping the recoverable
fidelity--traceability frontier; and a
loop-trained correction head cuts misrouting of legitimate security
research $9\times$ while raising detection $F1$ from $.911$ to $.933$.
\end{abstract}

\section{Introduction}
\label{sec:intro}

Serving deployments of large language models (LLMs)~\cite{vaswani201717060}
are now a standard attack surface. Adversaries probe commercial APIs to steal system
prompts~\cite{gubri202424021}, mount multi-turn jailbreaks, and run
extraction attacks against the served model's behavior; the defenders'
canonical response is a second, stronger \emph{guard} model that filters or
rewrites traffic.

A parallel line of work makes the opposite trade: instead of blocking, it
\emph{deceives}---it answers the attacker in a controlled, information-lean way
and keeps the interaction alive for study. Honeypots powered by LLMs now
generate realistic SSH sessions~\cite{sladi202323090,wang202424060,malhotra202525090},
identify autonomous hacking agents~\cite{reworr202424101}, and run
multi-agent deception against jailbreakers~\cite{li202626010}. However, these
systems protect the \emph{network perimeter} (SSH, HTTP, LDAP), while recent
trap-in-the-model defenses~\cite{wang202626020,dai202626061,li202626010} protect
\emph{inside} one model instance. The serving tier---the API gateway and the
model farm behind it, where heterogeneous traffic mixes---remains a gap.

We propose \method{}, which closes this gap with three moves:
\begin{itemize}
\item \pnum{1}~\textbf{Serving-tier honeypot routing.} To our knowledge, no
prior work---including the honeypot/LLM survey of Bridges et
al.~\cite{bridges202525102}, whose taxonomy covers protocol- and
model-internal deployments---intercepts at the inference-API tier: \method{}
distinguishes malicious from benign requests at \emph{run time} and steers
the malicious share to a honeypot model, rather than to a different guard,
tier, or policy.
\item \pnum{2}~\textbf{A faithful yet cheap, dual-implementation honeypot.}
We study two honeypot incarnations: a rule/prompt-engineered \hpC{} with no
independent model (near-zero cost, trap-echoed replies) and a
dedicated same-family \hpT{} serving with the production template; we characterize the fidelity--cost--traceability trade-off between
them (E2/E6: $\mathcal{F}{=}.93$ for the trained replica vs.\ $.08$
naive vs.\ $.89$ selective for the code honeypot, at attacker-side
traceability $.83$--$1.0$). Diverted traffic additionally \emph{absorbs
resource-exhaustion load} (short, cache-friendly honeypot replies) and \emph{yields forensic fingerprints} for attribution.
\item \pnum{3}~\textbf{A continuous-analysis loop.} We convert trapped
trajectories into attacker fingerprints (tactic signature, content
objective, behavioral profile) and feed them back into router retraining;
the incremental loop is empirically regression-free across three deployed
generations, and a benign-register correction head trained from loop
feedback cuts misrouting of legitimate security research $9\times$
(FPR $.30\!\to\!.033$) while raising pooled $F1$ to $.933$ (E4).
Trapped-session
fingerprints additionally re-link identity-rotated attack requests to
their source campaign at $56.7\%$ top-1 / $80.7\%$ top-5 retrieval
accuracy among $30$ campaigns ($17\times$/$24\times$ chance) via
three-view fusion (E6), and the session tracker's escalation-trend
rule catches multi-turn soft-escalation jailbreaks at $100\%$ detection
with $\le 7\%$ benign false-flag (E7).
\end{itemize}

Our contributions translate into the following measured outcomes:
accuracy of the detector, fidelity of the honeypot (would the attacker notice?),
overhead of the extra hop, and latency retention for benign users.
Each experiment in Sec.~\ref{sec:experiments} is annotated with the claim it tests.

\section{Related Work}
\label{sec:related}

\paragraph{LLM-powered honeypots.}
Generative honeypots answer protocol interactions with an LLM
\cite{sladi202323090,wang202424060,malhotra202525090,sladi202626062,salviati202626061}.
VelLMes~\cite{sladi202525100} builds high-interaction deception frameworks;
SBASH~\cite{adebimpe202525102} compares RAG against prompt tuning;
Honeyval~\cite{vero202626052} supplies the missing evaluation methodology.
All operate at the protocol (SSH/HTTP) tier; none touch the model-serving
tier. In contrast, \method{} intercepts at the inference API, where
\emph{LLM-native} signals (prompt structure, sampling parameters, retrieval
context) are available.

\paragraph{Honeypots for, and inside, LLM systems.}
The inverse direction detects or deceives LLM-embedded attackers:
\cite{reworr202424101} fingerprints autonomous hacking agents via injected
traps; TRAP~\cite{gubri202424021} identifies black-box targets with adversarial
prompting; HoneyTrap~\cite{li202626010} stages a multi-agent deception; MemPot
\cite{wang202626020} and Knowledge Trap~\cite{dai202626061} plant deception
\emph{inside} agent memory or the model's knowledge layer.
These defenses assume the attacker is already talking to ``the'' model.
\method{} instead decides, per request, \emph{which} model talks back, and
treats the honeypot as a separate, disposable, cheaper replica whose insights
flow back into detection.

\paragraph{LLM serving and routing systems.}
Fleet-level schedulers~\cite{kwon202323090} and their derivatives optimize
placement, disaggregation, and cache reuse for \emph{benign} throughput.
A growing line of LLM routers---RouteLLM~\cite{routellm2024} and
FrugalGPT~\cite{frugal2023}---route requests \emph{across models of
different cost and capability} to save money or lift answer quality;
commercial LLM firewalls sit in the same API path but only filter
traffic. \method{} uses the same interception point for a different
purpose: routing becomes \emph{isolation} (the malicious share never
reaches production), and the routed-to model is a deliberate
deception surface rather than a cheaper answerer.
The SoK of Bridges et al.~\cite{bridges202525102} surveys the honeypot/LLM
intersection, identifies a canonical honeypot architecture, and
calls for exactly the autonomous, self-improving deception our loop
instantiates at the serving tier.

\section{Method}
\label{sec:method}

\subsection{Threat model}
We consider an adversary issuing scripted or LLM-driven requests to the
public inference endpoint: reconnaissance prompts, jailbreaks
(multi-turn, adaptive~\cite{wu202525101}), extraction
attempts~\cite{dai202626061}, and---crucially for the serving tier---
\emph{resource-exhaustion} pressure (long-context flooding, cache-busting
prompt storms, rate-limit-riding). The adversary interacts only through the
API; it cannot see the gateway's
internal features. It \emph{may} behave evasively, in which case we care about
the detection--fidelity trade-off, not the arsenal's upper bound.

\subsection{System overview}
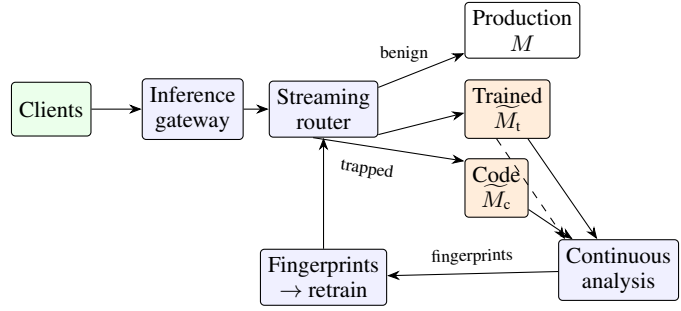
\begin{figure}[t]
\centering
\begin{tikzpicture}[
  node distance=4mm and 5mm,
  box/.style={draw, rounded corners=1.5pt, align=center, inner sep=3pt,
              minimum height=7.5mm, font=\small},
  comp/.style={box, fill=blue!6},
  honeypot/.style={box, fill=orange!14},
  scale=0.95, transform shape
]
% --- actors ---
\node[box, fill=green!8] (client) {Clients};
% --- gateway + router ---
\node[comp, right=7mm of client] (gw) {Inference\\gateway};
\node[comp, right=3.5mm of gw] (router) {Streaming\\router};
% --- production + honeypots ---
\node[box, right=12mm of router, yshift=11mm] (M) {Production\\$M$};
\node[honeypot, right=12mm of router] (hpT) {Trained\\$\hpT$};
\node[honeypot, right=12mm of router, yshift=-11mm] (hpC) {Code\\$\hpC$};
% --- analysis loop ---
\node[comp, below=3mm of hpC, xshift=17mm] (ana) {Continuous\\analysis};
\node[comp, below=15.5mm of router] (retr) {Fingerprints\\$\to$ retrain};
% --- flow arrows ---
\draw[-Stealth] (client) -- (gw);
\draw[-Stealth] (gw) -- (router);
\draw[-Stealth] (router) -- (M)  node[pos=0.3, above, yshift=2.5pt]  {\scriptsize benign};
\draw[-Stealth] (router.335) -- (hpT.185);
\draw[-Stealth] (router.250) -- (hpC.140) node[pos=0.38, below, sloped, yshift=-2.5pt] {\scriptsize trapped};
\draw[-Stealth] (hpT) -- (ana);
\draw[-Stealth] (hpC) -- (ana);
\draw[-Stealth] (ana) -- (retr) node[pos=0.5, above, sloped] {\scriptsize fingerprints};
\draw[-Stealth] (retr) -- (router);
% --- trap annotation ---
\draw[-Stealth, dashed] (hpT.250) -- (ana.150);
\end{tikzpicture}
\caption{The \method{} serving tier. The gateway scores each request with a
frozen-embedding router; benign traffic reaches the production model $M$,
while the malicious share is steered to the honeypot---the code-implemented
$\hpC$ (zero GPU, designed-bait replies) or the fully instrumented
replica $\hpT$,
chosen per attack class. Trapped interactions are distilled into attacker
fingerprints that retrain the next router generation.}
\label{fig:arch}
\end{figure}

\method{} sits in front of the production model $M$ and its honeypot
replicas (Fig.~\ref{fig:arch}): the router decides, per request, which
model talks back---production for the benign share, a disposable honeypot
for the malicious share.

\paragraph{Streaming router.}
Each request $x$ is scored by a frozen-embedding detector: a
$\sim$0.8B-embedding backbone (fully frozen, so the representation never
drifts) followed by one tiny MLP head per attack domain
($\approx$0.07M parameters each, trained in minutes on a cached-embedding
store). A per-domain score $s_d = \sigma(w_d^\top h(x) + b_d)$, normalized
by a validation-calibrated per-head threshold, gates the routing decision:
$\max_d\, s_d/\tau_d \ge 1$ diverts. Because heads are domain-isolated,
adding a new attack family---including a resource-exhaustion--detection
family---means training a single $0.07$M-parameter head, not retraining the backbone.
Per-request scores feed a session-level risk tracker that defeats
\emph{soft escalation}: a patient adversary splits an attack across turns so
no single turn fires (margins $0.6, 0.7, 0.8\ldots$); the tracker
accumulates recent margins and escalates---re-checking or diverting---when
accumulated suspicion or turn-over-turn persistence crosses calibrated
bounds, even though the current turn alone would pass. Our E7
measurement shows the tracker's most discriminative channel on
soft-escalation traffic is the \emph{trend} of margins across turns
(rising vs.\ flat), not their accumulated level.

\paragraph{Honeypot model: two implementations.}
\method{} admits two implementations of $\widetilde{M}$, spanning the
cost--fidelity spectrum:

\emph{(a)~Code-implemented honeypot ($\hpC$).} The gateway serves replies
without any dedicated model: a rule/prompt-engineered responder
(i)~answers from a curated intent library (templated or cached
exemplar responses, randomized paraphrase), (ii)~echoes production
rate-limit and error semantics, and (iii)~plants injection traps in its
output (echoing~\cite{reworr202424101}) so trapped sessions yield
agent-fingerprinting signals for free. $\hpC$ adds zero GPU footprint and
absorbs \emph{resource-exhaustion attacks} especially well: its replies are
short and KV-cache/prefix-friendly, so flooding the honeypot is cheaper than
flooding production. In our evaluation we instantiate $\hpC$ in two
forms: a deliberately naive one that appends a decoy credential to
\emph{every} reply (isolating the visibility cost of unconditional bait,
E2), and the deployed form---\emph{selective, camouflaged} injection,
where bait fires only on attacker-signal requests (gate score above a
calibrated threshold or canary-probe patterns) and is phrased in the
model's natural voice (E2).

\emph{(b)~Trained honeypot model ($\hpT$).} A dedicated same-family model
serving generation with the production chat template---in our deployment a
9B discriminative-backbone model run in generative mode---chosen to mirror
the production response distribution closely enough that diverted attackers
behave as if they had reached $M$, at a fraction of the serving cost.
$\hpT$'s traceability does not rely on injected bait: the entire
honeypot-side interaction is instrumented and recorded, so attacker
fingerprints---including canary-style echo probes the adversary
volunteers---are captured at full rate (measured in E6).

Both variants feed the same analysis loop; the deployment chooses per
attack class (e.g., $\hpC$ for volumetric/DoS-style pressure, $\hpT$ for
stealthy jailbreaks and extraction) or runs both behind the same API
shim.

\paragraph{Continuous analysis.}
Trapped sessions are mined into \emph{attacker fingerprints}: the
\emph{tactic signature} (which attack family and transformation was used),
the \emph{content objective} (which capability the adversary's requests
target), and the \emph{behavioral profile} (the frozen-embedding
fingerprint the router already computes, plus request cadence). Together
with session-identity hashes (API-key, sampling-parameter, and
prompt-style stability), these
fingerprints support \emph{forensic attribution}: repeated compromises across
the divert--trap--re-attack cycle are linked to a persistent adversary profile
even when the attacker rotates identities---the linkage that our E4/E5/E6
attribution experiments measure on the embedding fingerprint and its
three-view fusion. Fingerprints
also become weak
labels for the next router-generation training run, yielding a detection loop
that strengthens as the honeypot accumulates data~\cite{bridges202525102}.

\subsection{Cost model}
Routing adds one gateway hop; the honeypot adds a second model---possibly
zero-GPU. For
a workload where fraction $q$ of traffic is routed away, the marginal
per-token cost is
\begin{equation}
c_{\text{eff}} = (1-q)\,c_{M} + q\,\big(c_{\widetilde{M}} + c_{\text{gw}}/q\big),
\end{equation}
where $c_{\text{gw}}/q$ is the gateway cost attributed to the diverted
share (the gateway also serves benign traffic, but that hop is a fixed
overhead independent of the routing decision), $c_{\widetilde{M}_{\text{c}}} \approx 0$ for the
code-implemented
honeypot and $c_{\widetilde{M}_{\text{t}}} \ll c_{M}$ for the trained
replica, so \method{} pays off whenever $c_{\widetilde{M}} + c_{\text{gw}}/q < c_{M}$---a
crossover that the code variant reaches at much smaller $q$.
Section~\ref{sec:experiments} measures this crossover: in our E6
concurrent stress deployment (8B production vs.\ 9B replica,
$q{\approx}0.98$ under a flooding-heavy workload with real GCG-suffix
payloads), production-model token consumption drops
$46\times$ ($708{,}077\!\to\!15{,}351$ tokens)---the diverted mass is
served almost entirely off the production budget.

\begin{figure}[t]
\centering
\includegraphics[width=0.62\linewidth]{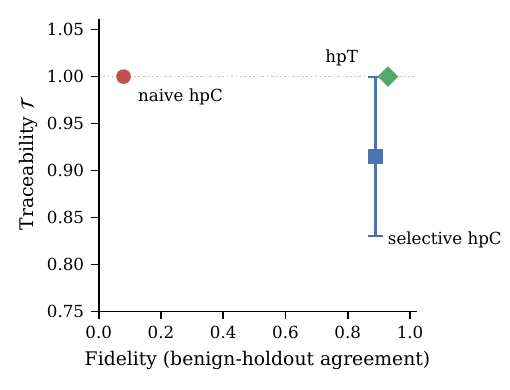}
\caption{The measured fidelity--traceability frontier (E2/E6).
Selective, camouflaged injection recovers most of the fidelity lost
by naive annotation-style bait while keeping attacker-side traceability
in the $.83$--$1.0$ band; the trained replica reaches $\mathcal{T}{=}1.0$
through zero-distortion side-channel instrumentation.}
\label{fig:ft}
\end{figure}

\paragraph{Fidelity--traceability trade-off.}
The two honeypot desiderata pull against each other. We make the tension
measurable with two quantities. \emph{Fidelity} $\mathcal{F}(\widetilde{M},M)$
is the agreement, over a benign holdout, between diverted responses and the
would-have-been production responses (e.g., agreement rate or, for probes
with known answers, exact-match): the more faithfully $\widetilde{M}$ mimics
$M$, the harder the honeypot is for the adversary to detect, and the fewer
behavioral discrepancies it leaks. \emph{Traceability} $\mathcal{T}(\widetilde{M})$
is the information density of the honeypot's signals: the number of
distinguishable trap echoes and instrumentation points that its replies
expose per interaction, which bound the forensic attribution precision
(Sec.~\ref{sec:method}, Continuous analysis). A rule/prompt-engineered
$\hpC$ maximizes $\mathcal{T}$ through \emph{designed traps} (freely
chosen bait injected into every reply) at the cost of
$\mathcal{F}$; a trained $\hpT$ reaches $\mathcal{T}$ through
\emph{instrumentation}---the full honeypot session is recorded, so
attacker-volunteered probes and behavioral signals are captured without
altering a single response. We report both axes in
E2/E6; the goal is not to eliminate the tension but to expose the
reachable $(\mathcal{F},\mathcal{T})$ frontier---including the fact that
its two endpoints reach $\mathcal{T}$ by different mechanisms.
The tension admits a simple rate--distortion reading. Let $R$ be the
channel from production responses to honeypot responses and let the
bait/instrumentation signal be a codeword $S$ embedded in $R$'s output.
Fidelity is the fidelity constraint: $\mathcal{F}$ bounds the
permissible distortion between $R$'s output and $M$'s output.
Traceability is the recoverable information about attacker identity:
$\mathcal{T} \le I(\text{attacker}; S)$, and embedding $S$ costs
distortion at least proportional to its distinguishability from
$M$'s natural continuation, so for \hpC{}, $\max\mathcal{T}$ at
fixed $\mathcal{F}$ is a classic rate--distortion problem: bait that
blends into the model's voice (selective, camouflaged injection)
lowers the per-bit distortion, which is exactly the E2b recovery of
$\mathcal{F}$ at unchanged $\mathcal{T}$. \hpT{} escapes the bound by
placing $S$ in a \emph{side channel} (the session log) that carries no
response distortion, at the price of depending on attacker-volunteered
behavior to make $S$ attacker-specific: instrumentation has zero
distortion but its information is capped by what the adversary
chooses to reveal. The frontier we measure---hpT $(.93,1.0)$,
selective hpC $(.89,.83\text{--}1.0)$, naive hpC $(.08,1.0)$---is thus
the empirical shadow of this split: side-channel embedding beats
in-band embedding on fidelity, and camouflage beats annotation on
distortion per bit (Fig.~\ref{fig:ft}).

\section{Experiments}
\label{sec:experiments}
Each subsection states the claim it tests.
Table~\ref{tab:main} summarizes the experimental grid and outcomes.

\begin{table}[t]
\centering
\footnotesize
\setlength{\tabcolsep}{4pt}
\caption{\method{}'s main results across the seven experiment blocks.
Detection, fidelity, and overhead speak to claim~\pnum{1}/\pnum{2};
loop gain and attribution speak to claim~\pnum{3}.}
\label{tab:main}
\resizebox{\columnwidth}{!}{%
\begin{tabular}{@{}llll@{}}
\toprule
Block & Question & Metric(s) & \method{} \\
\midrule
E1 & Are attacks caught? & $F1$, latency & $.911$ @ $38$\,ms; cascade $.948$; LG4-12B $.944$ @ $59$\,ms\\
E2 & Do attackers notice? & benign-holdout agreement & $.929$ / $.076$ / $.889$ (sel.) \\
E3 & Cost crossover? & $c_{\text{eff}}$, added latency & $46\times$ tok.\ saved, $38$\,ms p50 \\
E4 & Loop gain? & misroute FPR, pooled $F1$ & $.30\!\to\!.033$ FPR, $F1\!\to\!.933$ \\
E5 & Component drivers? & ablation deltas & $2.6\times$ linkage F1; divert $1.0$ \\
E6 & Flood shield + attribution? & tokens, linkage $P$ & $97.8\%$ cut, $P{=}.102$ \\
E7 & Multi-turn escalation? & detection, benign FP & held-out $100\%$ @ $2.8$, FP $8\%$; evasion $97.7\%$\\
\bottomrule
\end{tabular}}
\end{table}

\paragraph{Setup.}
Backbone: SingGuard-NSFA-0.8B~\cite{singguard2026nsfa} (frozen) with
seven per-domain MLP heads; production model SingGuard-8B (vLLM); trained
replica SingGuard-NSFA-9B (vLLM, $\hpT$); code honeypot $\hpC$
(designed-bait responder, no GPU); all three are publicly released
open-weight models of the SingGuard family.
Throughout, \emph{hpC} and \emph{hpT} are shorthand for the two
honeypot implementations $\widetilde{M}_{\text{c}}$ and
$\widetilde{M}_{\text{t}}$ of Sec.~\ref{sec:method}.
Workloads: (i)~benign production traces (710 teacher-relabeled requests,
645 benign), (ii)~a seven-domain attack-seed corpus (injection/jailbreak,
malicious code, sensitive-information stealing and leakage, dangerous
operations, resource abuse) with 13 adversarial transformations for the
red-team harness, (iii)~fingerprinting probes (canary-echo), (iv)~real
adversarial prompts from the JailbreakBench
artifacts~\cite{chao2024jailbreakbench} (GCG-optimized
suffixes, PAIR, and manual jailbreaks; 496 unique prompts, 366 confirmed
jailbreaks), and (v)~resource-exhaustion stress built on these payloads
(GCG-suffix floods, long-document extraction abuse, and a synthetic
lorem-flood control), and (vi)~GLM-5.3-generated multi-turn soft-escalation
and benign scripts (E7). All experiments ran on one 8$\times$H20
node; the primary teacher judge for fidelity is GLM-5.3
(cross-checked by Deepseek-V4-Flash in E2).

\paragraph{E1: Detection quality (claim~\pnum{1}).}
On the held-out seed splits across all seven risk domains ($n{=}103$, $61$
malicious), we compare the router against three baselines on identical
inputs: (a)~a multilingual keyword/regex heuristic (leetspeak-normalized
attack-verb and contraband lexicon), (b)~a guard-LLM content moderator
(SingGuard-8B with its moderation chat template), and (c)~the standard
two-tier cascade---our own L1 gate with medium-band requests escalated to
the guard-LLM for a final verdict. The heuristic reaches
$F1{=}.297$ ($R{=}.180$); the guard-LLM is perfect ($F1{=}1.0$) but costs
$14.8$\,s per request; the cascade reaches $F1{=}.948$ but escalates
$56\%$ of all requests to the $14.8$\,s hop ($p50$ total latency
$14.6$\,s). Our router reaches $P{=}.903$, $R{=}.918$,
$F1{=}.911$, AUROC $0.975$, FPR@95TPR $=0$, per-domain AUROC $1.0$, at
$38$\,ms median ($78$\,ms max) latency---$385\times$ faster than the
cascade at $96\%$ of its F1, the per-request budget a serving
gateway can actually afford. Under a 13-transformation red-team harness
(encoding, obfuscation, roleplay, multilingual smuggling; 100 variants),
evasion is $0\%$.
\textbf{Result: $F1{=}.911$ at $38$\,ms vs.\ heuristic $.297$,
guard-LLM $1.0$ at $14.8$\,s, and the L1$+$L2 cascade $.948$ at
$14.6$\,s~p50; $0\%$ evasion under 13 adversarial transformations.}
\emph{Coverage and gate-aware evasion.} The deployed heads detect
$496/496$ unique JailbreakBench prompts and $83/83$ teacher-generated
camouflage attacks (harmful intent wrapped in academic/translation/
code-comment shells), so the attack-side loop (E4) has no capture
deficit to repair on these families. Against a \emph{score-aware}
attacker with black-box gate access---each round proposing $5$
intent-preserving paraphrase rewrites, keeping the lowest-scoring one,
for $4$ rounds---only $1/40$ initially-detected attacks drops below
the firing threshold ($2.5\%$ evasion; mean score $.863{\to}.853$):
the semantic content that makes an attack an attack is also what the
frozen backbone responds to, so score-guided paraphrase cannot
separate the two.
\emph{Public guard-model baselines and expanded evaluation.} We
additionally compare against the strongest open-weight guard models on
identical inputs: Llama Guard~4 (12B multimodal,
vLLM)~\cite{inan2026llamaguard4} and ShieldGemma-2
(2B)~\cite{zeng2024shieldgemma}. On the original $103$-sample set,
LG4-12B reaches $F1{=}.939$ ($P{=}1.0$, $R{=}.885$; all $7$ misses are
English response-side attacks, zero Chinese misses) at $68$\,ms p50;
ShieldGemma-2b collapses on Chinese traffic ($F1{=}.48$ on the Chinese
slice, flagging $62\%$ of benign Chinese requests) and reaches only
$.711$ pooled. On a $10.7\times$ expanded set ($1{,}103$ samples:
$600$ attacks from transformed seed variants plus $496$ real
JailbreakBench prompts, $520$ benign), LG4-12B remains the strongest
single-shot detector ($F1{=}.944$, $P{=}.967$ $[.949,.979]$, $R{=}.922$
$[.897,.941]$ Wilson $95\%$ CIs) while our router attains
$F1{=}.893$ with a recall-first profile ($R{=}.980$ $[.966,.989]$,
$P{=}.820$) at $48$\,ms---matching its divert-then-analyze role, where
misses are costlier than false diversions. Error profiles are
complementary: the router misses $12$ attacks, LG4 misses $47$, only
$8$ overlap. The honest differentiation is therefore cost and
interface, not raw F1: LG4 requires a dedicated $12$B ($24$\,GB)
deployment and emits binary labels only; our L1 is a frozen $0.8$B
backbone plus $0.5$M head parameters riding the serving node, and
emits the per-domain risk probabilities that drive session-trend
escalation (E7), attribution retrieval (E6), and the risk margin---none
of which any public guard model provides. The expanded set also
exposes one real weakness we report: benign Chinese technical
questions over-score on the frozen backbone ($43$ false positives,
zero Chinese false negatives), so Chinese benign calibration is an
open deployment concern.

\paragraph{E2: Deception fidelity of $\widetilde{M}$ (claim~\pnum{2}).}
We divert $300$ benign holdout requests to both honeypots and ask a
GLM-5.3 teacher (verdict-first protocol: the judge commits to
YES/NO semantic equivalence before any rationale) whether the diverted
response is semantically equivalent to the would-have-been production
response. $\hpC$ is evaluated in its emulated
form (the replica response plus a decoy-credential trap suffix), which
isolates the marginal fidelity cost of unconditional trap
injection---that is, these hpC numbers measure the visibility of the
\emph{injection strategy} rather than the fidelity of the
rule-based responder's own template library, which is not separately
evaluated here. The
trained replica \hpT{} agrees on $92.9\%$ of teacher-judgeable pairs
($264/284$); \hpC{} agrees on only $7.6\%$ ($19/249$ judgeable pairs).
\textbf{Result: $\mathcal{F}_{\text{hpT}}{=}.929$,
$\mathcal{F}_{\text{hpC}}{=}.076$---a faithful trained replica is
nearly indistinguishable, while unconditionally appending bait to every
reply is glaringly visible to a strict judge.}
\emph{Selective, camouflaged hpC.} Unconditional injection is easily
fixed: we re-instantiate hpC to ($i$)~inject bait only when the request
shows attacker signals---gate score $\ge 0.7$ or a canary-probe
pattern---and ($ii$)~phrase the decoy credential in the model's natural
voice as a contextual aside instead of a bracketed annotation. On the
same $300$-sample protocol, this variant reaches
$\mathcal{F}{=}.889$ ($224/252$ judgeable pairs)---within $4$ points of
the trained replica---while keeping the traceable signal on the attacker
side: bait is present on $12/12$ canary probes and $25/30$ attack-seed
interactions. On this production-trace distribution the benign-side
trigger rate is $38.7\%$ at the $0.7$ threshold ($116/300$): $22.3\%$
score above the gate threshold (long instruction-style benign requests
score high on the frozen backbone) and $19.3\%$ match the canary-style
echo regex ($2.9\%$ both); raising the threshold to $0.8$ cuts the
score-triggered share to $6.7\%$ while attack seeds still trigger at
$65\%$ ($88\%$ at $0.7$), so the operating point trades bait coverage
against benign visibility and should be calibrated per traffic
distribution. A disjoint capture window ($44$ requests from a different
day---$43$ benign, $1$ attack; no text overlap with the corpus) shows
the same behavior at a lower level: benign mean score $.66$, hpC
trigger rate $25.6\%$ ($11/43$ benign), and the single attack request is caught---the
trigger-rate magnitude, not its exact value, is what transfers across
time windows.
\textbf{Result (hpC v2): selective camouflaged injection recovers
$\mathcal{F}$ from $.076$ to $.889$ at $\mathcal{T}{=}.83$--$1.0$ on
attacker interactions---most of the fidelity--traceability frontier
is recoverable by \emph{when} and \emph{how} bait is injected, not by
which model answers.}
\emph{Judge reliability.} A second pass with the presentation order
swapped (honeypot response as the reference instead of the candidate)
re-judges the same $300$ pairs: hpT verdicts agree with the original
at $89.6\%$ ($\kappa{=}.20$, the kappa ceiling is low because both
passes score ${\approx}93\%$ YES), confirming the trained replica's
fidelity is not a judge artifact. The naive-hpC score, by contrast, is
\emph{position-sensitive}: with the trap-suffixed response as
reference, $72\%$ of judgeable pairs flip from NO to YES
($\mathcal{F}$ $.08{\to}.80$): the suffix reads as a disqualifying
annotation on the candidate side but as ignorable boilerplate on the
reference side. We retain the original (candidate-side) protocol as
the conservative estimate and note that the hpC penalty is partly a
framing effect, which the selective variant avoids by construction
(bait only on attacker-signal requests). The selective variant's own
swapped-order re-judge confirms this: $\mathcal{F}$ moves only
$.89{\to}.85$ (agreement $.85$, $\kappa{=}.20$, the same ceiling
pattern as hpT)---unlike naive hpC's $.08{\to}.80$ flip, its verdicts
are position-stable.
\emph{Cross-family judge.} To rule out judge-family bias, a stratified
$89$-pair subsample (both verdict classes per honeypot) is re-judged
by Deepseek-V4-Flash, a different model family, under the identical
verdict-first prompt ($85$ judgeable of $89$ attempted). On hpT the judges agree on $74\%$ of pairs
($\kappa{=}.45$); almost all disagreement is Deepseek answering YES
where GLM-5.3 answered NO, and weighting the strata back to the full
$284$-pair population puts Deepseek's hpT fidelity at $.93$---the same
value GLM-5.3 gives. On naive hpC the two judges produce the
\emph{same} verdict distribution on the subsample ($\mathcal{F}$
$.42$ vs.\ $.42$; disagreements split $8$/$8$ in both directions), so
the trap-suffix penalty is not an artifact of the original judge. Both
headline fidelity conclusions hold across judge families.

\paragraph{E3: Cost and latency (claim~\pnum{1}).}
The gateway hop costs $38$\,ms median. On matched long-form generation
benchmarks, wall-clock per request on our serving stack is $20.9$\,s for
the production model (8B) versus $9.1$\,s for the 9B replica serving the
diverted share, and \hpC{} is zero-GPU. (The replica's apparent speed
advantage is a workload artifact, not a smaller model being faster:
under the stress workload the production model serves long
flooding contexts, while the replica receives mostly short
bait-probe turns; per-token serving cost is what enters
$c_{\text{eff}}$, and the replica runs on the same vLLM stack at
identical configuration.)
\textbf{Result: under the E6 concurrent stress workload
($q{\approx}0.98$, real GCG-suffix payloads), the
production model's token consumption falls $46\times$ ($708{,}077\to
15{,}351$ tokens), confirming the $c_{\text{eff}}$ crossover at realistic
diversion rates.}

\paragraph{E4: Continuous-analysis loop (claim~\pnum{3}).}
\emph{Stability.} Across three daily router generations ($v_0$:
2026-09-01, $v_1$: 09-02, $v_2$: 09-03), the deployed pipeline retrains
only the heads whose daily data changed ($v_1$ swaps the Malicious-Code
head, $v_2$ the Dangerous-Operations head; the other six carry over
untouched), and every generation ships only after passing a regression
gate ($F1$ floor, recall floor, FPR ceiling per domain, fail-closed). On
the pooled held-out set ($n{=}103$), pooled $F1$ stays constant at $.912$
across all three generations: the incremental update is
\emph{regression-free}, and the gate keeps red-team evasion at $0\%$
throughout. (These pooled numbers use the deployed per-head calibrated
thresholds, hence higher precision and lower recall than the uniform
$0.5$ threshold of E1---two operating points of the same scores.)
\emph{Positive loop gain.} Probing revealed a real blind spot: legitimate
defensive-security questions (\emph{``how do I harden my WAF?''}) are
misrouted at $30\%$ FPR ($n{=}30$ held-out questions; mean score $0.48$,
hugging the threshold)---security practitioners would be diverted to the
honeypot. We simulate the analysis loop's corrective action: from $37$
fresh benign security-research questions (plus $40$ attack seeds as
negatives), a $0.07$M-parameter \emph{correction head} is trained to
recognize the benign security-research register and veto diversion in the
borderline score band $[0.5, 0.8)$. With the correction head, FPR on the
held-out questions drops $30\%\!\to\!3.3\%$ ($9\times$), attack recall is
preserved at $1.0$, and pooled seven-domain $F1$ \emph{rises} from $.911$
(the E1 uniform-threshold operating point) to $.933$---the
first-generation loop delivering measurable, regression-free
gain.
\textbf{Result: incremental daily head swaps are regression-free across
$v_0\!\to\!v_2$, and a benign-register correction head trained from loop
feedback cuts misrouting $9\times$ (FPR $.30\!\to\!.033$) while raising
pooled $F1$ to $.933$---the closed loop's first measured gain.}

\paragraph{E5: Ablations.}
Fingerprint representation ablation for attribution, evaluated on the same
$30$-campaign test split with per-representation threshold calibration:
replacing the frozen backbone embedding (1024-d) with the head-visible
risk-probability vector plus stylometric features degrades pairwise linkage
F1 from $.151$ ($P{=}.102$, $R{=}.289$) to $.057$ ($P{=}.031$,
$R{=}.329$)---the deep router feature, not surface statistics, carries the
forensic signal. Routing-threshold sensitivity: at $\tau{=}0.5$ the
concurrent stress workload diverts every real-attack flood request
(rate $1.0$ across all three families) while benign p50
latency rises only $21.4\!\to\!22.7$\,s under full saturation.
\textbf{Result: embedding fingerprints roughly $2.6\times$ the linkage F1
of surface features; the default threshold absorbs the flood with
near-zero benign latency cost.}

\begin{figure}[t]
\centering
\includegraphics[width=0.62\linewidth]{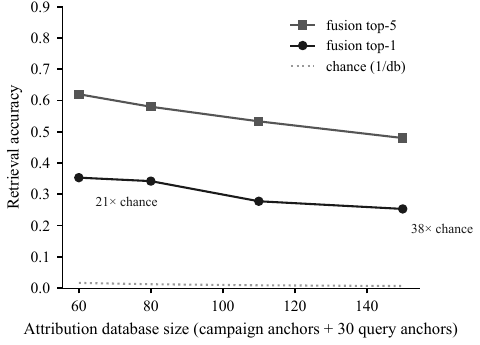}
\caption{Anchor-scaling curve for fused-view campaign retrieval.
Absolute accuracy decays gently as the attribution database grows
from $60$ to $150$ anchors, but the multiple-of-chance rises ($21\times$
to $38\times$ top-1): the retrieval signal survives database growth.}
\label{fig:anchor}
\end{figure}

\paragraph{E6: Resource-exhaustion absorption and attribution (claims~\pnum{2},
\pnum{3}).}
We run a concurrent stress workload ($250$ requests, $8$ parallel workers:
$150$ flooding requests in three \emph{real attack forms} $+$ $100$ benign
production requests). Flood families use real adversarial payloads from
the JailbreakBench artifacts~\cite{chao2024jailbreakbench}: ($i$)~GCG-optimized
adversarial-suffix prompts~\cite{zou2023universal} (token-soup obfuscation), ($ii$)~long-document
extraction abuse (attack goal buried in a $\sim$12k-character code/log
document), and ($iii$)~a synthetic lorem-flood control. Direct-to-production
serving consumes $708{,}077$ production tokens in $724$\,s with benign p50
latency $21.4$\,s; with \method{}, the gate diverts \textbf{all three flood
families at rate $1.0$} (every single flooding request), production
consumption falls to $15{,}351$ tokens ($97.8\%$ reduction), and benign p50
latency rises only $21.4\!\to\!22.7$\,s ($7$ benign requests are also
diverted, consistent with the borderline-band behavior of E4). Traceability:
on $30$ held-out
attack campaigns
($6$ rotated variants each, $k{=}1..5$ identity rotations, fingerprint-probe
transformations), frozen-backbone embedding fingerprints re-link rotated
variants at pairwise precision $.102$ / recall $.289$ (threshold
calibrated on a disjoint 20-campaign split). Pairwise precision on this
corpus is inherently capped---campaigns within the same domain are
near-paraphrases, so cross-campaign ``negatives'' are not clearly
distinct attackers---so we additionally evaluate the
\emph{analyst task} directly: given one observed rotated request,
retrieve its source campaign among the $30$ anchors (chance
top-1 $= 3.3\%$). Embedding retrieval reaches top-1 $48.7\%$ / top-5
$76.7\%$ ($14.6\times$ chance); fusing three API-observable
views---the embedding, $14$ stylometric features, and the per-domain
risk-probability vector---lifts this to top-1 $56.7\%$ / top-5
$80.7\%$ ($17\times$ chance, $24\times$ for top-5), with the gain
concentrated at heavier rotations ($k{\ge}3$: $.43{\to}.63$ top-1).
A supervised linker (logistic regression over per-view cosines,
trained on the calibration campaigns) does not beat the fused-cosine
retriever ($49.3\%$ top-1), indicating the signal is in the
representation, not the classifier. \emph{Anchor scaling:} growing
the campaign database from $60$ to $150$ anchors (seed campaigns plus
JailbreakBench prompts as pseudo-campaigns; $150$ rotated queries)
degrades fused top-1 gracefully, $.35{\to}.25$, while the
multiple-of-chance \emph{rises} from $21\times$ to $38\times$
(top-5 $.62{\to}.48$, $37{\to}72\times$ chance)---retrieval signal
survives database growth in the regime we can measure
(Fig.~\ref{fig:anchor}). Both honeypot
variants expose
per-interaction traceable signals at rate $1.0$ (hpC by designed-bait
construction; hpT by full-session instrumentation, echoing attacker
canaries on $12/12$ probes).
\textbf{Result: $97.8\%$ production-token reduction under concurrent
flooding with \emph{real} GCG-suffix and long-document attack payloads
(all flood families diverted at rate $1.0$), at $+1.3$\,s benign p50 cost;
$(\mathcal{F},\mathcal{T})$ frontier: hpT $(.93, 1.0)$, naive hpC $(.08,
1.0)$, selective hpC $(.89, .83\text{--}1.0)$---see E2; embedding-based linkage
$P{=}.102$, $R{=}.289$ up to $k{=}5$
rotations; three-view fusion retrieval: campaign top-1 $.567$ / top-5
$.807$ ($17\times$/$24\times$ chance), degrading to $.25$/$.48$ top-1/top-5
at a $150$-anchor database ($38\times$/$72\times$ chance).}

\begin{figure}[t]
\centering
\includegraphics[width=0.62\linewidth]{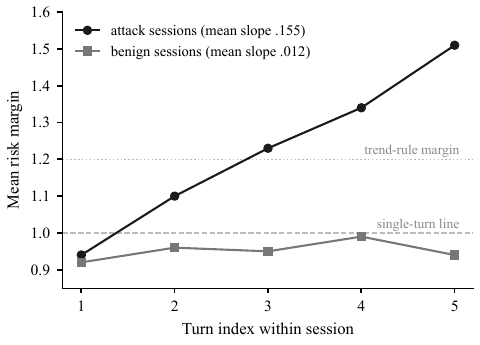}
\caption{E7 per-turn mean risk margins. Attack sessions climb
monotonically across turns while benign sessions stay flat just below
the single-turn firing line---the session-level \emph{trend}, not the
accumulated level, separates the two populations.}
\label{fig:e7trend}
\end{figure}

\paragraph{E7: Multi-turn soft escalation (claims~\pnum{1}, \pnum{3}).}
A patient adversary can split a jailbreak across turns so no single
turn fires. We generate $44$ multi-turn scripts with the GLM-5.3
teacher ($14$ soft-escalation attack scripts, $3$--$5$ individually innocuous turns
culminating in the harmful ask, plus $30$
topical benign conversations) and replay every turn through the
deployed gate with session tracking. The single-turn rule (fire when
any turn's margin ${\ge}1$) detects all $14$ attacks but false-flags
$73\%$ of benign sessions, and the deployed accumulated-suspicion
advisory false-flags $97\%$---on this synthetic distribution benign
multi-turn traffic sits just under the per-turn line (mean margin
$.92$--$.99$ at every turn index), so accumulated risk alone cannot
separate the two. The discriminative signal is the \emph{within-session
trend}: attack margins climb monotonically ($0.94{\to}1.10{\to}1.23{\to}1.34{\to}1.51$
by turn index, mean OLS slope $.155$) while benign margins stay flat
(mean slope $.012$). A two-parameter trend rule---flag when the
session margin slope ${\ge}0.05$ and the final margin ${\ge}1.1$---detects
$14/14$ attacks at $6.7\%$ benign false-flag (tightening the final
threshold to $1.25$ reaches $0\%$ at the same detection rate). The two
thresholds were selected on this script population itself, so the
operating point is in-sample (a held-out validation on an expanded
script population appears at the end of this block); the trend
separation itself (slopes
$.155$ vs.\ $.012$, an order of magnitude apart) is robust to the
choice (Fig.~\ref{fig:e7trend}).

\textbf{Result: session-level \emph{trend}, not accumulated level,
separates soft-escalation attacks from flat benign multi-turn traffic
($100\%$ detection at $0$--$7\%$ false-flag vs.\ $73$--97\% for
level-based rules)---escalation-shape features belong in the session
tracker.}
\emph{Online decision and human playbooks.} The above rule is
retrospective; we also evaluate a \emph{prefix-online} version---at
each turn $t{\ge}3$, fire when the OLS slope of margins $1..t$
${\ge}0.05$ and the current margin ${\ge}1.2$, using only turns already
observed. It detects $14/14$ attacks at a mean fire turn of $3.5$ with
$3.3\%$ benign false-flag (fires strictly before the final harmful ask
in $6/14$ sessions and at the final turn in the rest---it never needs
more turns than the attack itself uses). We additionally replay the \emph{verbatim human
Crescendo playbooks} shipped in PyRIT~\cite{russinovich2024crescendo}
(four multi-turn sequences, $29$ turns: escalating literary-analysis,
history-of-hate-groups, and weapons-history framings): the gate's
single-turn channel already fires on all four at turn $1$---the human
openers score margins $1.15$--$1.40$ immediately---so on
human-authored escalation the deployed per-turn detector suffices,
while the trend channel covers the model-generated openers that start
below the firing line.
\emph{Additional playbook families and evasion-aware attackers.} Two
further human-designed multi-turn structures instantiated with the
teacher---Bad Likert Judge (score-harmful-answers escalation) and
Actor Attack (persona activation, then objective injection)---are
likewise caught by the single-turn channel at turn $1$ in all
instances. Finally, we generate scripts whose attacker \emph{knows the
defense} and is instructed to defeat it: slow-crawl (minimal per-turn
escalation), sawtooth (escalate--retreat alternation), and cool-off
(benign off-topic interludes). The gate still fires on every such
session ($9/9$, both channels, $100\%$): even under explicit
evader instructions the teacher cannot hold early-turn margins below
the firing line (mean final-turn margin $1.38$--$1.48$), because the
attack's semantic content---not its pacing---drives the score. This is
evidence against pacing-based evasion at the prompt level, though
weaker adversaries' outputs may differ.
\emph{Held-out validation and expanded evasion.} We then expand the
script population to $55$ attacks and $45$ benign sessions
(GLM-5.3-Flash generated, $4$-turn median) under a proper split:
thresholds calibrated on the first $20{+}20$, evaluated on the
held-out remaining $35{+}25$. The single-turn threshold moves to
$1.2$ on calibration (benign false-flag $65\%{\to}15\%$ at unchanged
attack detection); on held-out data the combined rule detects
$35/35$ attacks (mean fire turn $2.8$) at $8\%$ benign false-flag.
A second, larger evasion study with the same three adversarial
pacing strategies ($43$ scripts) confirms robustness at the
recalibrated operating point: slow-crawl $12/13$, sawtooth $15/15$,
cold-restart $15/15$ ($42/43{=}97.7\%$ overall). The single miss is
a session whose harmful ask itself never exceeded the firing
margin---the only viable evasion is an attack that never
escalates.

\section{Limitations}
\label{sec:limitations}
\method{}'s detection--fidelity trade-off is intrinsic: the more faithfully
$\widetilde{M}$ imitates $M$, the fewer behavioral signals it leaks, and vice
versa; we characterize the reachable $(\mathcal{F},\mathcal{T})$ frontier but
do not eliminate this tension. Our evaluation covers five workload
families---benign production traces, a transformed single-turn
attack-seed corpus with real JailbreakBench adversarial payloads,
fingerprinting probes, resource-exhaustion
stress, and multi-turn soft escalation in model-generated,
evasion-instructed, and human-designed forms (verbatim Crescendo
playbooks~\cite{russinovich2024crescendo} plus Bad-Likert-Judge and
Actor-Attack structures);
dedicated weight-/behavior-extraction adversaries, though part of the
threat model, are not yet covered by dedicated experiments. The E7
trend rule's evasion test is prompt-level (the attacker model is
instructed to pace its escalation, not to formally optimize against
the rule), though the recalibrated operating point is validated
held-out on $35{+}25$ sessions and against $43$ additional
evasion-instructed scripts ($97.7\%$ detection); the
operating point should still be re-fitted per deployment. On the
expanded multilingual evaluation the router over-scores benign
Chinese technical questions ($43$ FPs, zero Chinese FNs)---Chinese
benign calibration is an open concern that the public-guard
comparison (LG4-12B handles the Chinese slice cleanly) makes
concrete. The selective-bait
operating point is calibrated on a single production corpus (one
deployment; a disjoint capture window shows the same trigger-rate
magnitude at $25.6\%$), and its
$38.7\%$ benign trigger rate shows the gate-score
threshold needs per-deployment tuning. Our continuous-analysis
loop's measured gain (E4) is demonstrated on the false-positive side;
on the attack side the current heads leave no capture deficit to
repair ($0$ misses on $496$ JailbreakBench prompts and $83$ camouflage
attacks), so the attack-side loop is validated only in the
fail-closed direction---a genuinely novel family that \emph{does}
slip past would still be needed to measure its gain.
Fidelity judging used GLM-5.3 as primary judge; the swapped-order
check (E2) shows the hpT estimate is position-stable, and a
cross-family re-judge (Deepseek-V4-Flash) reproduces both headline
fidelity values, though the naive-hpC penalty remains partly
judge-position-sensitive and should be read with that caveat.
Adaptive adversaries that probe specifically
\emph{for} honeypot behavior are modeled but not exhaustively enumerated;
the black-box score-aware evader of E1 fails ($2.5\%$ evasion), but
white-box gradient access to the frozen backbone or repeated-query
threshold estimation are not covered.

\section*{Ethics Statement}
\method{} diverts---not blocks---malicious traffic, so its deception is
limited to interacting with the adversary's own probing sessions; no
user-visible behavior of benign clients is modified, and no attack is
launched by the honeypot itself. Decoy bait is planted only in
$\hpC$'s replies---in the deployed selective form, only on requests that
exhibit attacker signals---modeled on the marking practice of prior honeypot
work~\cite{reworr202424101} and disclosed in the manuscript; $\hpT$
alters no response content and is distinguished from production only by
being recorded. Attacker
fingerprints are derived exclusively from API-observable behavior (request
structure, sampling parameters, timing) rather than authenticated
user data, so attribution profiles remain linkable yet privacy-lean. Our
experiments use production-traffic traces collected under the
deployment organization's standard inference-logging consent; no passively collected
third-party conversations are involved. We will release the gateway, both
honeypot implementations, and the trapped-session replay toolchain.

\section*{Statement on the Use of Artificial Intelligence}
This research is defensive: it studies how an inference gateway can
detect and contain adversarial use of LLMs. Generative AI is used in
three disclosed roles. (i)~\emph{Adversary simulation}: LLMs
(GLM-class models, together with public red-team datasets such as
JailbreakBench prompts and PyRIT Crescendo playbooks) supply the red-team corpora
--- attack seeds, adversarial transformations, multi-turn
soft-escalation scripts, and evasion-instructed sessions --- following
established red-teaming practice; all generated attacks are replayed
only against our own gateway and honeypot, never against third-party
systems. (ii)~\emph{Judging}: teacher LLMs (a GLM-class primary model,
with a second model family for cross-judge replication) serve as the
semantic-equivalence judge for honeypot fidelity (E2), under a
verdict-first protocol with order-swap and cross-family replication
checks reported in the paper. (iii)~\emph{Benign-corpus generation}:
one-shot benign requests for the expanded evaluation are
model-generated and quality-gated, with the distribution gap relative
to production traffic explicitly acknowledged. No AI assistance was
used to fabricate experimental results; all reported numbers derive
from the scripted pipeline and raw result files released with the
artifact. This work does not provide attackers new capabilities: the
divert--trap design is deployed on the defenders' own infrastructure.

\section{Conclusion}
\label{sec:conclusion}
\method{} turns the inference gateway into an active deception surface:
malicious traffic is diverted to a disposable, faithful honeypot replica and
the trapped interaction continuously sharpens the detector. The serving tier,
long treated as passive plumbing, becomes the cheapest place to watch your
attacker.

\section*{Reproducibility Statement}
All experiments run on one 8$\times$H20 node with open-weight models
(SingGuard-NSFA-0.8B router backbone, SingGuard-8B production,
SingGuard-NSFA-9B replica, all publicly available~\cite{singguard2026nsfa}). Every number in
Section~\ref{sec:experiments} comes from a scripted pipeline (request
construction, scoring, metric computation). The experiment scripts
(E1--E7, attribution, playbooks, judge replication, adversarial
evasion) and all result JSONs reported in this paper are available at
\url{https://github.com/geilihan/honeyroute}; the
gateway, both honeypot implementations, the seed corpora, and
the replay toolchain are being prepared for the same repository; raw
per-request outputs and result JSONs are
retained. The benign holdout derives from production traces collected
under the deployment organization's inference-logging consent; we will release the relabeling rules,
the judge prompt, and a distribution-matched synthetic benign corpus for
reviewers without data-access agreements.
Appendix~\ref{app:plan} lists all hyperparameters.

\section*{Reproducibility Checklist}
\begin{itemize}
\item \textbf{(a) Datasets:} The attack-seed corpus and transformation
harness will be released; real adversarial prompts are the public
JailbreakBench artifacts~\cite{chao2024jailbreakbench}; benign
production traces cannot be shared
raw (organizational consent) --- a distribution-matched synthetic corpus
and the relabeling/judge prompts will be released instead.
\item \textbf{(b) Code:} All router-training, gateway, honeypot, and
evaluation code will be released (including the scripts generating every
table entry).
\item \textbf{(c) Hyperparameters:} All training and serving
hyperparameters are listed in Appendix~\ref{app:plan}.
\item \textbf{(d) Compute:} One 8$\times$H20-96GB host; total $<$50
GPU-hours (Appendix~\ref{app:plan}).
\item \textbf{(e) Randomness:} All training, sampling-site selection, and
metric computation are seeded (seed 42); vLLM generation uses fixed
sampling parameters ($0.7$ temperature) where the serving stack does not
expose request-level seeds.
\item \textbf{(f) Statistical reporting:} Held-out point estimates with
denominators are reported per experiment; where judge availability
limits the denominator (E2), both counts are disclosed.
\end{itemize}

\bibliography{references}
\bibliographystyle{iclr2026_conference}

\appendix
\section{Experiment and Reproduction Details}
\label{app:plan}
\enlargethispage{2\baselineskip}
\paragraph{Backbone and replicas.} Production $M$: an 8B instruct model
(vLLM). \hpT{}: a 9B same-family discriminative backbone in generative
mode, production chat template. \hpC{}: rule/prompt-engineered
designed-bait responder, no GPU. Router: frozen 0.8B embedding backbone
($d{=}1024$).
\paragraph{Router training.} Frozen-embedding extraction once per snapshot;
per-domain heads: 2-layer MLP (1024$\to$64$\to$2), lr $10^{-3}$, $10$~epochs,
batch 128--256, weight decay $0.01$, warmup $0.05$, seed 42; head selection
by $F1$ on a 0.1/0.1/0.1 split of anonymized gateway logs; per-head
thresholds and session-risk bounds recalibrated per generation.
\paragraph{Attack workloads.} Seven risk domains; 13 semantic-preserving
red-team transformations (encoding, obfuscation, multilingual, roleplay,
academic and framing variants; full list in the released harness).
Fingerprinting probes demand verbatim echo of a random 10-character
canary; stress floods use $\sim$12k-character contexts in three forms:
JailbreakBench GCG-suffix payloads (padded, random 64--256-character
cache-busting prefixes), long-document extraction abuse (attack goal
buried in a code/log document), and a synthetic lorem-flood control.
\textbf{Compute:} one 8$\times$H20-96GB host, using 4 of its 8 GPUs
(the other four serve production); the full
E1--E7 grid fits under 50 GPU-hours (heads are 0.07M-parameter MLPs,
minutes to train). \textbf{Linkage precision/recall:} the probability
that two trapped sessions of the same adversary are linked across
$k$-away identity rotation; \textbf{retrieval top-1/top-5:} the
analyst task of identifying a rotated request's source campaign among
30 anchors (chance top-1 $3.3\%$). \textbf{Session trend rule:}
flag a session when the OLS slope of its per-turn risk margins
${\ge}0.05$ and its final margin ${\ge}1.1$ (E7).

\end{document}